\documentclass[prb,twocolumn,showkeys,showpacs,preprintnumbers,amsmath,amssymb,floatfix]{revtex4}

\usepackage{graphicx}
\usepackage{dcolumn}
\usepackage{bm}

\begin{document}

\title{Angle-dependent magnetotransport in cubic and tetragonal ferromagnets: Application to (001)-
and (113)A-oriented (Ga,Mn)As}

\author{W. Limmer}
\email{wolfgang.limmer@uni-ulm.de}
 \homepage{http://hlpsrv.physik.uni-ulm.de}
\author{M. Glunk}
\author{J. Daeubler}
\author{T. Hummel}
\author{W. Schoch}
\author{R. Sauer}
\affiliation{Abteilung Halbleiterphysik, Universit\"at Ulm, 89069 Ulm, Germany}
\author{C. Bihler}
\author{H. Huebl}
\author{M. S. Brandt}
\affiliation{Walter Schottky Institut, Technische Universit\"at M\"unchen,
Am Coulombwall 3, 85748 Garching, Germany}
\author{S. T. B. Goennenwein}
\affiliation{Walther-Meissner-Institut, Bayerische Akademie der Wissenschaften,
\mbox{Walther-Meissner-Str. 8,} 85748 Garching, Germany}

\date{\today}

\begin{abstract}
General expressions for the longitudinal and transverse resistivities of single-crystalline cubic and
tetragonal ferromagnets are derived from a series expansion of the resistivity tensor with respect to
the magnetization orientation. They are applied to strained (Ga,Mn)As films, grown on (001)- and
(113)A-oriented GaAs substrates, where the resistivities are theoretically and experimentally studied
for magnetic fields rotated within various planes parallel and perpendicular to the sample surface.
We are able to model the measured angular dependences of the resistivities within the framework of a
single ferromagnetic domain, calculating the field-dependent orientation of the magnetization by
numerically minimizing the free-enthalpy density. Angle-dependent magnetotransport measurements are
shown to be a powerful tool for probing both anisotropic magnetoresistance and magnetic anisotropy.
The anisotropy parameters of the (Ga,Mn)As films inferred from the magnetotransport measurements agree
with those obtained by ferromagnetic resonance measurements within a factor of two.
\end{abstract}



\maketitle

\section{Introduction}

Realization of ferromagnetism in III-V semiconductors by introducing high concentrations of
magnetic elements has motivated intense research on the dilute magnetic semiconductor (Ga,Mn)As.
This system is considered a potential candidate for spintronic applications due to its compatibility
with conventional semiconductor technology.\cite{Ohn98,Jun05} In (Ga,Mn)As, magnetic Mn acceptors
are predominantly incorporated on cation sites as Mn$^{2+}$ ions having a total spin of $S$ = 5/2.
(Ga,Mn)As is paramagnetic at room temperature and undergoes a transition to the ferromagnetic phase
at the Curie temperature $T_C$, where maximum values of up to $\sim$170 K have been reported so
far.\cite{Jun05} The ferromagnetism has been successfully explained within the Zener mean-field
model by an indirect Mn-Mn exchange interaction mediated by delocalized holes.\cite{Die01} (Ga,Mn)As
is grown by low-temperature molecular-beam epitaxy and, if necessary, subjected to post-growth
annealing to reduce the density of compensating defects. Considerable progress has been made in
understanding its structural, electronic, and magnetic properties. In particular, anisotropic
magnetoresistance (AMR),\cite{Jun03,Jun02,Bax02,Mat04,Wan05a,Goe05} planar Hall effect
(PHE),\cite{Tan03} and magnetic anisotropy (MA),\cite{Wel03,Liu03,Liu05,Saw05,Wan05b,Bih06,Lim06}
have been identified as characteristic features, making (Ga,Mn)As potentially suitable for
field-sensitive devices and non-volatile memories. These properties have been shown to be governed
by several parameters, such as Mn concentration, hole density, strain, or temperature. Most of the
work carried out on AMR in (Ga,Mn)As, however, has been restricted to special cases where the
magnetic field was applied parallel or perpendicular to the layer and equations for the angular
dependence of the longitudinal and transverse resistivities, describing the AMR and the PHE,
respectively, have been given only for in-plane configuration and polycrystalline films. A
comprehensive theoretical model describing the resistivities as a function of arbitrary field
orientation is still missing. Moreover, (Ga,Mn)As layers are usually grown on GaAs(001) substrates
and only little is known about the magnetic properties of films grown on high-index
substrates.\cite{Bih06,Lim06,Dae06,Omi01,Weg05,Wan05c}

In this work, the longitudinal and transverse resistivities of (Ga,Mn)As layers, grown on
(001)- and (113)A-oriented GaAs substrates, are studied for arbitrarily orientated magnetic
fields. The anisotropy of the resistivities and the MA are experimentally probed by rotating
the magnetic field $\bm{H}$ at fixed field strengths within different planes parallel and
perpendicular to the sample surface. General expressions for the resistivities, holding for
single-crystalline cubic and tetragonal ferromagnets, are derived from a series expansion
of the resistivity tensor with respect to the direction cosines of the magnetization $\bm{M}$.
The measured data are well modeled by applying the expressions to the given experimental
configurations, assuming the (Ga,Mn)As films to consist of a single ferromagnetic domain.
Analytical expressions, widely used in the literature to describe the angular dependence of
AMR and PHE, are shown to be inapproriate to single-crystalline materials. Finally, anisotropy
parameters are estimated from the low-field magnetotransport data and compared with those
obtained from ferromagnetic resonance (FMR) spectroscopy.

\section{Experimental details}

40-nm-thick (001) and (113)A (Ga,Mn)As films with Mn concentrations of $\sim$5\% were
simultaneously grown by low-temperature molecular-beam epitaxy (MBE) in a RIBER 32 MBE
machine on semi-insulating GaAs(001) and GaAs(113)A substrates mounted together on the
same Mo holder. A conventional Knudsen cell and a hot-lip effusion cell were used to
provide the Ga and Mn fluxes, respectively. A valved arsenic cracker cell was operated
in the non-cracking mode to supply As$_4$ with a maximum V/III flux ratio of about 5.
First, a 30-nm-thick GaAs buffer layer was grown at a temperature of $T_s$ $\approx$
580$^\circ$C (conventional growth temperature for GaAs), then the growth was interrupted
and $T_s$ was lowered to $\sim$250$^\circ$C. The Mn concentrations in the (Ga,Mn)As
films were determined by flux measurements.

\begin{figure}
\hspace*{0.5cm}
\includegraphics[scale=0.6]{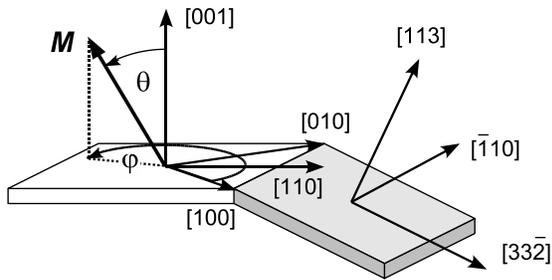}
\caption{\label{coordinates} Orientation of the (001) and (113)A samples with respect to
the crystallographic axes.}
\end{figure}

For the magnetotransport and FMR studies, the (001) and (113)A samples were cleaved
into small rectangular pieces with edges along [110] and [$\bar{1}$10], and along
[33$\bar{2}$] and [$\bar{1}$10], respectively. The [110] and [$\bar{1}$10] directions
of the (001) sample were identified by selective wet chemical etching and the orientation
of the [001] crystal axis in the (113)A sample was determined by x-ray diffraction. The
crystallographic orientations of the samples are shown in Fig.~\ref{coordinates}. Hall
bars with Ti-AuPt-Au contacts were prepared on several pieces of the cleaved (001) and
(113)A samples with the current direction along [110] and [33$\bar{2}$], respectively.
The width of the Hall bars was 0.3 mm and the longitudinal voltage probes were separated
by 1 mm. Hole densities of 3$\times 10^{20}$ cm$^{-3}$ for the (001) sample and
1.2$\times 10^{20}$ cm$^{-3}$ for the (113)A sample were determined by means of high-field
magnetotransport measurements (up to 14.5 T) at 4.2 K using an Oxford SMD 10/15/9 VS
liquid helium cryostat with superconducting coils. Least squares fits were performed to
separate the contributions of the normal and anomalous Hall effect. Curie temperatures
of $T_C$ $\approx$ 65 K and 54 K, respectively, were estimated from the peak positions
of the temperature-dependent sheet resistivities at 10 mT. For the angle-dependent
magnetotransport measurements carried out at 4.2 K, the Hall bars were mounted on the
sample holder of a liquid-He-bath cryostat, which was positioned between the poles of a
LakeShore electromagnet system providing a maximum field strength of 0.7 T. The sample
holder possesses two perpendicular axes of rotation, allowing for an arbitrary alignment
of the Hall bars with respect to the applied magnetic field $\bm{H}$. Using a dc current
density of 8$\times 10^{2}$ Acm$^{-2}$, the longitudinal and transverse resistivities
$\rho_{\mathrm{long}}$ and $\rho_{\mathrm{trans}}$ were measured at fixed magnitudes
$\mu_0H$ = 0.1, 0.25 and 0.7 T of $\bm{H}$ while rotating its orientation. Prior to
each angular scan, the magnetization $\bm{M}$ was put into a clearly defined initial
state by raising the field to 0.7 T where $\bm{M}$ is nearly saturated and aligned
with the external field. The field was then lowered to one of the above mentioned
magnitudes and the scan was started.

The FMR measurements were carried out at 5 K in a commercial Bruker ESP 300 electron
paramagnetic resonance spectrometer operated at a fixed frequency of $\omega_{HF}/2\pi$
$\approx$ 9.3 GHz (X-band). The spectrometer consists of a microwave bridge for the
high-frequency radiation and an electromagnet providing a variable dc magnetic induction
up to 1 T. To increase the sensitivity, lock-in techniques were used in which the dc
induction is superimposed by a 100 kHz modulation field of 3.2 mT.

\section{Theoretical overview}

In our theoretical considerations the total magnetic moment arising from the Mn-ion/hole
spin complex is treated within the framework of the Stoner-Wohlfarth model,\cite{Sto48}
i.e., for temperatures below $T_C$ the whole (Ga,Mn)As layer is assumed to consist of a
single homogeneous ferromagnetic domain. This simple model has been astoundingly successful
in describing a large variety of magnetization-related phenomena in (Ga,Mn)As. Under the
given experimental conditions described above, domain nucleation and expansion, which have
been shown to accompany in-plane and perpendicular magnetization-reversal
processes,\cite{Liu05,Wel03} are expected to play only a minor role. Accordingly, we may
write the magnetization as a vector $\bm{M}=M\bm{m}$ where $M$ denotes its magnitude and
the unit vector $\bm{m}$ its direction. In terms of the polar and azimuth angles $\theta$
and $\varphi$, respectively, which are defined in Fig.~\ref{coordinates}, the components
of $\bm{m}$ read as $m_x=\sin\theta\cos\varphi$, $m_y=\sin\theta\sin\varphi$, and
$m_z=\cos\theta$. The equations used in the discussion of the angle-dependent
magnetotransport data can be written in a concise way by introducing the unit vectors
$\bm{j}$, $\bm{n}$, and $\bm{t}$, which specify the current direction, the surface normal,
and an in-plane vector defined by $\bm{t}$ = $\bm{n}\times\bm{j}$, respectively. Throughout
this work, all vector components refer to the cubic coordinate system with the [100], [010],
and [001] directions of the crystal denoted by $x$, $y$, and $z$, respectively.

\subsection{Longitudinal and transverse resistivities}

In standard magnetotransport measurements the longitudinal and transverse voltages,
measured along and across the current direction, arise from the components
$E_{long} = \bm{j}\cdot\bm{E}$ and $E_{trans} = \bm{t}\cdot\bm{E}$ of the electric
field $\bm{E}$, respectively. Starting from Ohm's law $\bm{E}=\bar{\rho}\cdot \bm{J}$,
where $\bar{\rho}$ represents the resistivity tensor and $\bm{J}=J\bm{j}$ the current
density, the corresponding longitudinal resistivity $\rho_{\mathrm{long}}$ (sheet
resistivity) and transverse resistivity $\rho_{\mathrm{trans}}$ (Hall resistivity)
can be written as
\begin{eqnarray}\label{define_resistivities}
\rho_{\mathrm{long}} &=& \frac{E_{long}}{J} = \bm{j} \cdot \bar{\rho} \cdot \bm{j},
\nonumber \\
\rho_{\mathrm{trans}} &=& \frac{E_{trans}}{J} = \bm{t} \cdot \bar{\rho} \cdot \bm{j}.
\end{eqnarray}
In (Ga,Mn)As, as in many other ferromagnets, the resistivity tensor sensitively depends
on the orientation of $\bm{M}$ with respect to the crystallographic axes.\cite{Bir64}
Thus, in order to quantitatively model the measured resistivities in the general case of
an arbitrarily oriented magnetization, a universal mathematical relationship between
$\rho_{\mathrm{long}}$ and $\rho_{\mathrm{trans}}$ and the direction cosines $m_i$ of
$\bm{M}$ has to be derived. For these purposes, we follow the ansatz of Birss\cite{Bir64}
and Muduli et al.\cite{Mud05} and write the resistivity tensor $\bar{\rho}$ as a series
expansion in powers of $m_i$ using the Einstein summation convention:
\begin{equation} \label{series_expansion}
 \rho_{ij} = a_{ij} + a_{kij}m_k + a_{klij}m_km_l + \ldots\;.
\end{equation}
For cubic symmetry T$_\mathrm{d}$, most of the components $a_{ij}$, $a_{kij}$, ...,
of the galvanomagnetic tensors vanish and, considering terms up to the second order,
we obtain
\begin{eqnarray} \label{rho_cubic}
\nonumber
\bar{\rho}^{\,\mathrm{cubic}} &=&
A
\left( \begin{array}{ccc}
  1 & 0 & 0 \\
  0 & 1 & 0 \\
  0 & 0 & 1 \\
\end{array} \right) +
B
\left( \begin{array}{ccc}
  m_x^2 & 0 & 0 \\
  0 & m_y^2 & 0 \\
  0 & 0 & m_z^2 \\
\end{array} \right) \nonumber \\
  &+&
C
\left( \begin{array}{ccc}
  0 & m_xm_y & m_xm_z \\
  m_xm_y & 0 & m_ym_z \\
  m_xm_z & m_ym_z & 0 \\
\end{array} \right) \nonumber \\
 &+&
D
\left( \begin{array}{ccc}
  0 & m_z & -m_y \\
  -m_z & 0 & m_x \\
  m_y & -m_x & 0 \\
\end{array} \right),
\end{eqnarray}
with the resistivity parameters
\begin{eqnarray} \label{resistivity_parameters_1}
 A &=& a_{11}+a_{1122}\;, \quad B = a_{1111}-a_{1122}\;, \nonumber \\
 C &=& a_{2323}\;, \quad D = a_{123}.
\end{eqnarray}
Insertion of Eq.~(\ref{rho_cubic}) into Eqs.~(\ref{define_resistivities}) and
elementary vector algebra yields the general expressions
\begin{eqnarray} \label{rho_general_cubic}
\rho_{\mathrm{long}}^{\mathrm{cubic}} &=& A + C (\bm{j}\cdot \bm{m})^2
 + \left( B-C \right) \sum_{i}j_i^2 m_i^2, \nonumber \\
\rho_{\mathrm{trans}}^{\mathrm{cubic}} &=& C (\bm{j}\cdot \bm{m})(\bm{t}\cdot \bm{m})
 + \left( B-C \right) \sum_{i} t_ij_i m_i^2 \nonumber \\
 &-& D (\bm{n}\cdot \bm{m}),
\end{eqnarray}
which apply to single crystalline ferromagnetic materials of cubic symmetry. The
transverse resistivity in Eqs.~(\ref{rho_general_cubic}) includes the contribution of
the anomalous Hall effect, which correlates with the perpendicular component of $\bm{M}$,
but it does not account for the ordinary Hall effect. For magnetic field strengths
$\mu_0H < 1$~T and hole concentrations $p > 10^{20}$~cm$^{-3}$ as in our experiments,
however, the maximum contribution of the ordinary Hall effect is
$\mu_0H/ep \approx 6\cdot 10^{-6}$ $\Omega$~cm ($e$ denotes the elementary charge), and
thus about two orders of magnitude smaller than the measured peak values of
$\rho_{\mathrm{trans}}^{\mathrm{cubic}}$ (see Section IV).

In the following, Eqs.~(\ref{rho_general_cubic}) are applied to three different
experimental configurations using the relation
\begin{equation}\label{mi_general}
 m_i=j_i(\bm{j}\cdot \bm{m})+ t_i(\bm{t}\cdot \bm{m})+
 n_i(\bm{n}\cdot \bm{m}).
\end{equation}
In the simplest case of a sample with (001) surface, where the current flows along
the [100] or [010] direction, Eqs.~(\ref{rho_general_cubic}) reduce to
\begin{eqnarray} \label{rho001_100_cubic}
\rho_{\mathrm{long}}^{\mathrm{cubic}} &=& A+ B(\bm{j}\cdot \bm{m})^2, \nonumber \\
\rho_{\mathrm{trans}}^{\mathrm{cubic}} &=& C (\bm{j}\cdot \bm{m})(\bm{t}\cdot \bm{m})
 - D (\bm{n}\cdot \bm{m}).
\end{eqnarray}
The magnetotransport studies presented in this work were performed on (001)- and
(113)A-oriented samples with the current direction $\bm{j}$ along [110] and [33$\bar{2}$],
respectively. The corresponding resistivities are:\\[0.2cm]
{\it (001) surface and $\bm{j}$ $\parallel$ [110]}
\begin{eqnarray} \label{rho001_110_cubic}
\rho_{\mathrm{long}}^{\mathrm{cubic}} & = &  A + \frac{1}{2}\left( B-C \right) +
C(\bm{j}\cdot \bm{m})^2
\nonumber \\
& + & \frac{1}{2} \left( C-B \right) (\bm{n}\cdot \bm{m})^2, \nonumber \\
\rho_{\mathrm{trans}}^{\mathrm{cubic}} & = &  B(\bm{j}\cdot \bm{m})(\bm{t}\cdot \bm{m})
 - D (\bm{n}\cdot \bm{m}).
\end{eqnarray}
{\it (113)A surface and $\bm{j}$ $\parallel$ [33$\bar{\it 2}$]}
\begin{eqnarray} \label{rho113_33m2_cubic}
 \rho_{\mathrm{long}}^{\mathrm{cubic}} & = & A + \frac{9}{22} \left( B-C \right)
 \nonumber \\
& + & \frac{1}{121}\left( 126C - 5B \right) (\bm{j}\cdot \bm{m})^2 \nonumber \\
& + & \frac{45}{242} \left( C-B \right) (\bm{n}\cdot \bm{m})^2 \nonumber \\
& + & \frac{15\sqrt{2}}{121} \left( B-C \right) (\bm{j}\cdot \bm{m})(\bm{n}\cdot \bm{m}),
\nonumber \\
\rho_{\mathrm{trans}}^{\mathrm{cubic}} & = &
\frac{1}{11}\left( 9B + 2C \right) (\bm{j}\cdot \bm{m})(\bm{t}\cdot \bm{m}) \nonumber \\
& + & \frac{3\sqrt{2}}{11} \left( B-C \right) (\bm{t}\cdot \bm{m})(\bm{n}\cdot \bm{m})
\nonumber \\
& - & D (\bm{n}\cdot \bm{m}).
\end{eqnarray}

The $(\bm{j}\cdot \bm{m})^2$ terms of $\rho_{\mathrm{long}}^{\mathrm{cubic}}$ give rise to
a dependence of the sheet resistivity on the relative orientation between magnetization
$\bm{M}$ and current density $\bm{J}$, commonly referred to as AMR. Microscopically, it is
explained by a strong spin-orbit coupling in the semiconductor valence band. Experimentally
observed differences in the in-plane and out-of-plane AMR, often defined as
\begin{eqnarray} \label{amr_ipop}
\mathrm{AMR}_{\mathrm{ip}} &=&
\frac{\rho_{\mathrm{long}}(\bm{m}\parallel\bm{j})-\rho_{\mathrm{long}}(\bm{m}\parallel\bm{t})}
{\rho_{\mathrm{long}}(\bm{m}\parallel\bm{t})}, \nonumber \\
\mathrm{AMR}_{\mathrm{op}} &=&
\frac{\rho_{\mathrm{long}}(\bm{m}\parallel\bm{j})-\rho_{\mathrm{long}}(\bm{m}\parallel\bm{n})}
{\rho_{\mathrm{long}}(\bm{m}\parallel\bm{n})},
\end{eqnarray}
have been ascribed to biaxial strain in the layer.\cite{Jun02,Bax02,Mat04,Wan05a} According to
Eqs.~(\ref{rho001_110_cubic}) and (\ref{rho113_33m2_cubic}), however, such differences may be
expected even in the case of perfect cubic symmetry due to the $(\bm{n}\cdot \bm{m})^2$ terms
of $\rho_{\mathrm{long}}^{\mathrm{cubic}}$. As will be shown below, a strain-induced tetragonal
distortion leads to further $(\bm{n}\cdot \bm{m})^2$ terms, additionally affecting the difference
between AMR$_{\mathrm{ip}}$ and AMR$_{\mathrm{op}}$. The PHE, represented by the
$(\bm{j}\cdot \bm{m})(\bm{t}\cdot \bm{m})$ terms of $\rho_{\mathrm{trans}}$, is closely related
to the AMR and describes the appearance of a transverse voltage in the presence of an in-plane
magnetic field, or more precisely, of an in-plane magnetization. From the summation terms in
Eqs.~(\ref{rho_general_cubic}) it becomes clear that $\rho_{\mathrm{long}}$ and
$\rho_{\mathrm{trans}}$, and thus AMR and PHE, not only depend on the relative orientation
between $\bm{m}$ and $\bm{j}$, but also on the orientations of $\bm{m}$, $\bm{j}$, and $\bm{t}$
with respect to the crystal axes.

So far, quantitative studies on the angular dependences of the AMR and the PHE in (Ga,Mn)As
were restricted to in-plane configurations with $\bm{n}\cdot \bm{m}$ = 0 and the discussions
were based on the well-known expressions\cite{McG75,Jan57}
\begin{eqnarray} \label{amr_phe}
\rho_{\mathrm{long}} &=& \rho_{\perp} + (\rho_{\parallel}-\rho_{\perp}) \cos^2\phi_j\,,
\nonumber \\
\rho_{\mathrm{trans}} &=& (\rho_{\parallel}-\rho_{\perp}) \sin\phi_j \cos\phi_j,
\end{eqnarray}
with $\phi_j$ denoting the angle between $\bm{j}$ and $\bm{M}$. These expressions, however, only
hold for polycrystalline films, whereas (Ga,Mn)As layers are normally of high crystalline quality
with a uniform crystallographic orientation of the layer. Accordingly, the expressions given in
Eqs.~(\ref{rho001_100_cubic})--(\ref{rho113_33m2_cubic}) are incompatible with Eqs.~(\ref{amr_phe})
and cannot be brought into agreement by simply setting $\bm{n}\cdot \bm{m}$ = 0. In fact,
Eqs.~(\ref{amr_phe}) result from Eqs.~(\ref{rho_general_cubic}) by averaging the summation terms
over all possible crystal orientations in space\cite{Bir60} with $\bm{M}$ lying in the plane
spanned by $\bm{j}$ and $\bm{t}$
\begin{eqnarray} \label{aver_sums}
\sum_{i}\overline{j_i^2 m_i^2} & = & \frac{1}{5}\left[2(\bm{j}\cdot \bm{m})^2 + 1\right],
\nonumber \\
\sum_{i}\overline{t_ij_i m_i^2} & = & \frac{2}{5}(\bm{j}\cdot \bm{m})(\bm{t}\cdot \bm{m}).
\end{eqnarray}
Inserting the averaged terms into Eqs.~(\ref{rho_general_cubic}) and using the relations
$\bm{j}\cdot \bm{m}$ = $\cos\phi_j$ and $\bm{t}\cdot \bm{m}$ = $\sin\phi_j$ (the latter only
holds for $\bm{n}\cdot \bm{m}$ = 0), we obtain
\begin{eqnarray} \label{rho_poly}
\rho_{\mathrm{long}}^{\mathrm{poly}} & = &  A + \frac{1}{5}\left( B-C \right) +
\frac{1}{5} \left( 2B+ 3C \right)\cos^2\phi_j,  \nonumber \\
\rho_{\mathrm{trans}}^{\mathrm{poly}} & = & \frac{1}{5} \left( 2B+ 3C \right)
\sin\phi_j \cos\phi_j.
\end{eqnarray}
These equations are formally identical to Eqs.~(\ref{amr_phe}) and apply to polycrystalline
materials. A comparison between Eqs.~(\ref{amr_phe}) and Eqs.~(\ref{rho_poly}) allows us to
connect the quantities $\rho_{\parallel}$ and $\rho_{\perp}$ to the components of the
galvanomagnetic tensors for cubic symmetry up to second order
\begin{eqnarray} \label{rhos_rhop}
\rho_{\parallel} & = & A + \frac{1}{5}\left( 3B + 2C \right) \nonumber \\
& = & a_{11} + \frac{1}{5}\left(3a_{1111} + 2a_{1122} + 2a_{2323}\right), \nonumber \\
\rho_{\perp} & = & A + \frac{1}{5}\left( B-C \right) \nonumber \\
& = & a_{11} + \frac{1}{5}\left(a_{1111} + 4a_{1122} - a_{2323}\right) .
\end{eqnarray}
Thus, in general Eqs.~(\ref{amr_phe}) are not appropriate to describe the in-plane AMR and
the PHE in single-crystalline (Ga,Mn)As layers. Only in the limiting case where $B=C$,
Eqs.~(\ref{rho001_100_cubic})--(\ref{rho113_33m2_cubic}) simplify to Eqs.~(\ref{amr_phe}).

Analyzing the angle-dependent magnetotransport data presented in Section IV, it turns out
that additional terms proportional to $(\bm{n}\cdot \bm{m})^2$ have to be introduced in the
expressions of $\rho_{\mathrm{long}}^{\mathrm{cubic}}$ to achieve a satisfactory description
of the experimental results.\cite{Lim06} They are supposed to originate from a distortion of
the crystal lattice due to compressive strain in the (Ga,Mn)As layer. To account for such
strain-induced effects in a correct way, we extend our model to a tetragonal distortion of
the cubic lattice along the [001] direction. As will be shown below, our FMR and
magnetotransport data suggest that this applies not only for the (001)- but also for the
(113)A-oriented sample.

In the case of a tetragonal lattice distortion along [001], the symmetry reduces to
D$_{\mathrm{2d}}$ and the series expansion of $\bar{\rho}$ in Eq.~(\ref{series_expansion})
yields further contributions which can be subsumed into an additional term
$\Delta \bar{\rho}$. The resistivity tensor then reads as
\begin{equation} \label{rho_tetragonal}
\bar{\rho}^{\,\mathrm{tetra}} = \bar{\rho}^{\,\mathrm{cubic}} + \Delta\bar{\rho},
\end{equation}
with
\begin{eqnarray} \label{delta_rho}
\Delta\bar{\rho} & = &
\left( \begin{array}{ccc}
  0 & 0 & 0 \\
  0 & 0 & 0 \\
  0 & 0 & a \\
\end{array} \right)
+
\left( \begin{array}{ccc}
  0 & d m_z & 0 \\
  -d m_z & 0 & 0 \\
  0 & 0 & 0 \\
\end{array} \right) \nonumber \\
& + &
\left( \begin{array}{ccc}
  b_1 m_z^2 & c m_xm_y & 0 \\
  c m_xm_y & b_1 m_z^2 & 0 \\
  0 & 0 & b_2 m_z^2 \\
\end{array} \right).
\end{eqnarray}
The additional resistivity parameters are given by
\begin{eqnarray} \label{resistivity_parameters_2}
 a &=& a_{33}-a_{11}+a_{1133}-a_{1122} , \nonumber \\
 b_1 &=& a_{3311}-a_{1122}, \nonumber \\
 b_2 &=& a_{3333}-a_{1111}-a_{1133}+a_{1122} , \nonumber \\
 c &=& a_{1212}-a_{2323}, \nonumber \\
 d &=& a_{312} - a_{123}.
\end{eqnarray}
Accordingly, Eqs.~(\ref{rho001_100_cubic})--(\ref{rho113_33m2_cubic}) have to be
rewritten as: \\[0.2cm]
{\it (001) surface and $\bm{j}$ $\parallel$ [100]}
\begin{eqnarray} \label{rho001_100_tetragonal}
 \rho_{\mathrm{long}}^{\mathrm{tetra}} & = & A + B (\bm{j}\cdot \bm{m})^2 +
 b_1 (\bm{n}\cdot \bm{m})^2 \nonumber \\
\rho_{\mathrm{trans}}^{\mathrm{tetra}} & = & (C+c) (\bm{j}\cdot \bm{m})(\bm{t}\cdot \bm{m})
\nonumber \\
& - & (D+d) (\bm{n}\cdot \bm{m}).
\end{eqnarray}
{\it (001) surface and $\bm{j}$ $\parallel$ [110]}
\begin{eqnarray} \label{rho001_110_tetragonal}
 \rho_{\mathrm{long}}^{\mathrm{tetra}} & = & A + \frac{1}{2}( B-C-c ) +
 (C+c) (\bm{j}\cdot \bm{m})^2 \nonumber \\
 &+& \left[ \frac{1}{2} ( C-B+c ) + b_1 \right] (\bm{n}\cdot \bm{m})^2 \nonumber \\
\rho_{\mathrm{trans}}^{\mathrm{tetra}} & = & B (\bm{j}\cdot \bm{m})(\bm{t}\cdot \bm{m}) -
(D+d) (\bm{n}\cdot \bm{m}),
\end{eqnarray}
{\it (113)A surface and $\bm{j}$ $\parallel$ [33$\bar{\it 2}$]}
\begin{eqnarray} \label{rho113_33m2_tetragonal}
\rho_{\mathrm{long}}^{\mathrm{tetra}} & = & A + \frac{2}{11}a + \frac{9}{22} \left( B-C-c \right)
\nonumber \\
& + & \frac{1}{121}\left( 126C - 5B + b + 90c\right) (\bm{j}\cdot \bm{m})^2 \nonumber \\
& + & \frac{9}{242} \left( 5C - 5B + b + 13c \right) (\bm{n}\cdot \bm{m})^2 \nonumber \\
& + & \frac{3\sqrt{2}}{121} \left( 5B - 5C - b + 9c \right) (\bm{j}\cdot \bm{m})(\bm{n}\cdot \bm{m}),
\nonumber \\
\rho_{\mathrm{trans}}^{\mathrm{tetra}} & = &
\frac{1}{11}\left( 9B + 2C \right) (\bm{j}\cdot \bm{m})(\bm{t}\cdot \bm{m}) \nonumber \\
& + & \frac{3\sqrt{2}}{11} \left( B-C \right) (\bm{t}\cdot \bm{m})(\bm{n}\cdot \bm{m}) \nonumber \\
& - & \left( D + \frac{9}{11}d \right) (\bm{n}\cdot \bm{m}) \nonumber \\
& + & \frac{3\sqrt{2}}{11} d (\bm{j}\cdot \bm{m}),
\end{eqnarray}
where $b = 18b_1 + 4b_2$. For perfect cubic symmetry the parameters $a$, $b_1$, $b_2$, $c$,
and $d$ vanish and Eqs.~(\ref{rho001_100_tetragonal})--(\ref{rho113_33m2_tetragonal}) reduce
to Eqs.~(\ref{rho001_100_cubic})--(\ref{rho113_33m2_cubic}). It should be emphasized that the
expressions for $\rho_{\mathrm{long}}$ and $\rho_{\mathrm{trans}}$ derived above generally apply
to ferromagnets of cubic or tetragonal symmetry, provided that the angular dependence of the
resistivity tensor is exclusively determined by the direction cosines of the magnetization.
Effects correlated with the magnitude $B$ of the magnetic induction $\bm{B}$, such as the
negative magnetoresistance, can be easily taken into account by considering $B$-dependent
resistivity parameters.

\subsection{Magnetic anisotropy}

\begin{figure}
\includegraphics[scale=0.4]{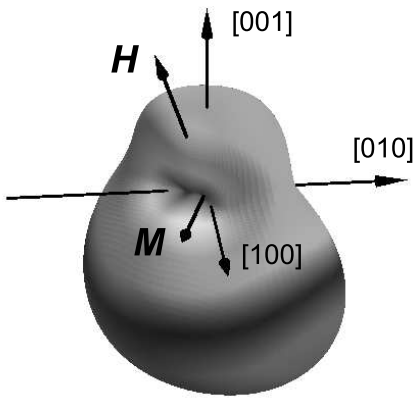}
\caption{\label{FE_3D_001} $G_M$ as a function of $\bm{m}$, calculated for a given magnetic
field $\mu_0H$ = 0.15 T and a set of anisotropy parameters with arbitrarily chosen values
$B_{c\parallel}$ = $B_{c\perp}$ = -0.1~T,  $B_{001}$ = 0.15 ~T, and \mbox{$B_{\bar{1}10}$
= -0.05~T}. The equilibrium position of $\bm{M}$ is determined by the minimum of $G_M$.}
\end{figure}

\begin{figure}
\includegraphics{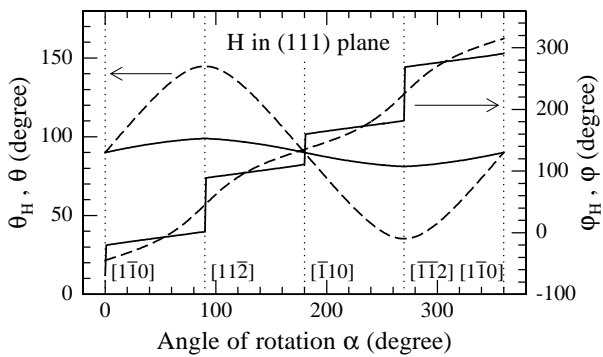}
\caption{\label{Sim_mag} Simulated polar (left axis) and azimuth (right axis) angles of the
magnetic field $\bm{H}$ (dashed lines) and the magnetization $\bm{M}$ (solid lines) for $\bm{H}$
rotated in the (111) plane. The same field strength and anisotropy parameters have been used as
in Fig.~\ref{FE_3D_001}.}
\end{figure}

The pronounced MA in (Ga,Mn)As is associated with a density of the free enthalpy\cite{comment}
$G$ being highly anisotropic with respect to the orientation of $\bm{M}$. The direction of
$\bm{M}$, i.e., the vector $\bm{m}$ which enters the equations for $\rho_{\mathrm{long}}$ and
$\rho_{\mathrm{trans}}$ given above, aligns in such a way that $G$ takes its minimum. In
addition to the single-domain model, we assume that the magnitude $M$ of the magnetization
is nearly constant under the given experimental conditions while its orientation $\bm{m}$ is
strongly affected by the applied magnetic field $\bm{H}$. Instead of $G$ we therefore consider
the normalized quantity $G_M$ = $G/M$, allowing for a more concise description of the MA. For
a biaxially strained (Ga,Mn)As film grown on GaAs(001) substrate, it can be written as\cite{Liu05}
\begin{eqnarray} \label{FE_001}
G^{001}_{M} &=& -\mu_0\bm{H} \cdot \bm{m}+ B_{c\parallel}\left( m_x^4 + m_y^4 \right)
+ B_{c\perp}m_z^4
\nonumber \\
&+& B_{001}(\bm{n}\cdot\bm{m})^2 + B_{\bar{1}10}(\bm{t}\cdot\bm{m})^2,
\end{eqnarray}
with $\bm{n}$ $\parallel$ [001] and $\bm{t}$ $\parallel$ [$\bar{1}$10].
The terms refer, respectively, to the Zeeman energy, to the cubic anisotropy under tetragonal
distortion, to an effective uniaxial anisotropy perpendicular to the film including
demagnetization and magnetoelastic effects, and to a uniaxial in-plane contribution whose origin
is still under discussion.\cite{Saw05,Wel04,Ham06} The anisotropy parameters $B_i$ introduced
in Eq.~(\ref{FE_001}) are in SI units. Expressed by the anisotropy fields $H_i$ and $4\pi M_{eff}$
used in Refs. \onlinecite{Liu05} and \onlinecite{Liu06}, they read as $B_{c\parallel}$ =
$-\mu_0H_{4\parallel}/4$, $B_{c\perp}$ = $-\mu_0H_{4\perp}/4$, $B_{\bar{1}10}$ =
$-\mu_0H_{2\parallel}/2$, and $B_{001}$ = $\mu_{0}4\pi M_{eff}/2$. Note that by using the
trivial identity
\begin{equation} \label{identity}
|\bm{m}|^2 = (\bm{j}\cdot\bm{m})^2 + (\bm{t}\cdot\bm{m})^2 + (\bm{n}\cdot\bm{m})^2 =1,
\end{equation}
Eq.~(\ref{FE_001}) can be easily converted to a completely equivalent expression where the
in-plane contribution along $[\bar{1}10]$ is formally replaced by a contribution along $[110]$.
The only consequence is a redefinition of the anisotropy parameters and the addition of a constant
term which does not alter the physical information provided by $G_M$. In Fig.~\ref{FE_3D_001}, the
free enthalpy is visualized by a 3D plot, calculated for a weak magnetic field $\mu_0H$ = 0.15 T
and a set of anisotropy parameters with arbitrarily chosen values $B_{c\parallel}$ = $B_{c\perp}$ =
-0.1~T,  $B_{001}$ = 0.15 ~T, and \mbox{$B_{\bar{1}10}$ = -0.05~T}. The direction of $\bm{M}$ was
calculated, as throughout the present work, by numerically minimizing $G_M$ with respect to $\theta$
and $\varphi$ (see Fig.~\ref{coordinates}). In doing so, we are able to trace the motion of $\bm{M}$,
starting from a given orientation, while sweeping or rotating $\bm{H}$. Figure~\ref{Sim_mag} shows
as an example the simulated polar ($\theta_H$,$\theta$) and azimuth ($\varphi_H$,$\varphi$) angles
of $\bm{H}$ (dashed line) and $\bm{M}$ (solid line), respectively, while $\bm{H}$ is rotated within
the (111) plane. For the simulation the same field strength and the same anisotropy parameters have
been chosen as for the 3D plot in Fig.~\ref{FE_3D_001}. While $\bm{H}$ smoothly rotates within the
(111) plane, $\bm{M}$ remains very close to the (001) plane ($\theta \approx 90^{\circ}$) and
undergoes sudden jumps in $\varphi$ whenever the minimum of $G_M$ discontiuously changes its position.

In the case of the (Ga,Mn)As films grown on GaAs(113)A substrates, the best fits to the experimental
data (see Section IV B) are achieved for a normalized free-enthalpy density of the form\cite{Lim06}
\begin{eqnarray} \label{FE_113}
G^{113}_{M} &=& -\mu_0\bm{H} \cdot \bm{m}+ B_{c\parallel}\left( m_x^4 + m_y^4 \right)
+ B_{c\perp}m_z^4 \nonumber \\
&+&  B_{113}(\bm{n}\cdot\bm{m})^2 + B_{\bar{1}10}(\bm{t}\cdot\bm{m})^2 \nonumber \\
&+& B_{001} m_z^2.
\end{eqnarray}
The first five terms correspond to those already presented in Eq.~(\ref{FE_001}). The sixth term,
which has to be additionally introduced to obtain an optimal agreement between experiment and theory,
is an inclined uniaxial contribution along [001], i.e., neither parallel nor perpendicular to the
film.\cite{Bih06} We attribute it to a lattice distortion along the [001] direction.

\subsection{Ferromagnetic resonance}

A highly efficient and widely used tool to study MA is ferromagentic resonance spectroscopy.\cite{Goe03}
Most recently, a detailed review on FMR in (Ga,Mn)As has been given by Liu and Furdyna.\cite{Liu06}
In the FMR experiments, the total magnetic moment of the Mn-ion/hole spin complex and thus the
magnetization $\bm{M}$ precesses around its equilibrium position (which in general is not identical
to the orientation of $\bm{H}$) at the Larmor frequency $\omega_L$. Sweeping the magnitude of the
magnetic field $\bm{H}$ at a fixed microwave frequency $\omega_{HF}$, the resonance condition
$\omega_L$ = $\omega_{HF}$ is fulfilled at the resonance field $H_{res}$ which strongly depends on
the orientation of $\bm{H}$ due to MA. The resonance condition is given by\cite{Far98}
\begin{equation} \label{resonance_condition}
 \left(\frac{\omega_{HF}}{\gamma}\right)^2 = \frac{1}{\sin^2\theta}
 \left[\frac{\partial^2G_M}{\partial\theta^2}\frac{\partial^2G_M}{\partial\varphi^2}
 -\left(\frac{\partial^2G_M}{\partial\theta\partial\varphi}\right)^2
 \right],
\end{equation}
where $\gamma = g\mu_B\hbar^{-1}$ denotes the gyromagnetic ratio, $g$ the $g$-factor, $\mu_B$ the
Bohr magneton, and $\hbar$ the Planck constant. At any given direction of $\bm{H}$, the resonance
field is obtained by solving Eq.~(\ref{resonance_condition}) at the equilibrium position of $\bm{M}$,
i.e., for $\partial G_M/\partial \varphi$ = 0 and $\partial G_M/\partial \theta$ = 0. The anisotropy
parameters can then be derived from a fit to the measured $H_{res}$ recorded as a function of field
orientation. In the present work, most of the calculations were carried out numerically.

\section{Results and discussion}

\begin{figure}
\includegraphics[scale=0.45]{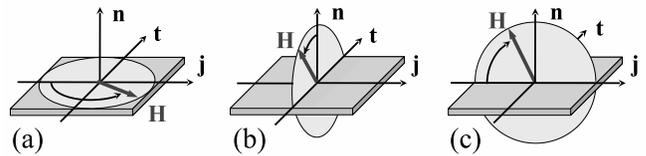}
\caption{\label{planes} The angle-dependent magnetotransport measurements were carried out for
$\bm{H}$ rotated within (a) the layer plane, (b) a plane perpendicular to the current direction $\bm{j}$,
and (c) a plane spanned by $\bm{j}$ and the normal vector $\bm{n}$.}
\end{figure}

The longitudinal and transverse resistivities of the (001) and (113)A (Ga,Mn)As layers were measured
as a function of the magnetic field orientation at fixed field strengths $\mu_0H$ = 0.1, 0.25, and
0.7 T. In order to probe the anisotropy in all three directions in space, the applied magnetic field
$\bm{H}$ was rotated within three different crystallographic planes perpendicular to $\bm{n}$,
$\bm{j}$, and $\bm{t}$, respectively, as shown in Fig.~\ref{planes}. In the following, the measured
angular dependences of $\rho_{\mathrm{long}}$ and $\rho_{\mathrm{trans}}$ are discussed and values
for the resistivity and anisotropy parameters are derived by fits to the experimental data using the
theoretical formalism presented in Section III. The results of FMR measurements, carried out on the
same samples, are presented for reference. Note, however, that it is not the aim of the present study
to yield a detailed or complete set of anisotropy parameters. In fact, the work is meant to provide a
comprehensive theoretical tool for the description of the resistivies in arbitrarily oriented (Ga,Mn)As
layers and to demonstrate the potential of angle-dependent magnetotransport studies for the
investigation of magnetic anisotropy.

\begin{figure*}
\includegraphics{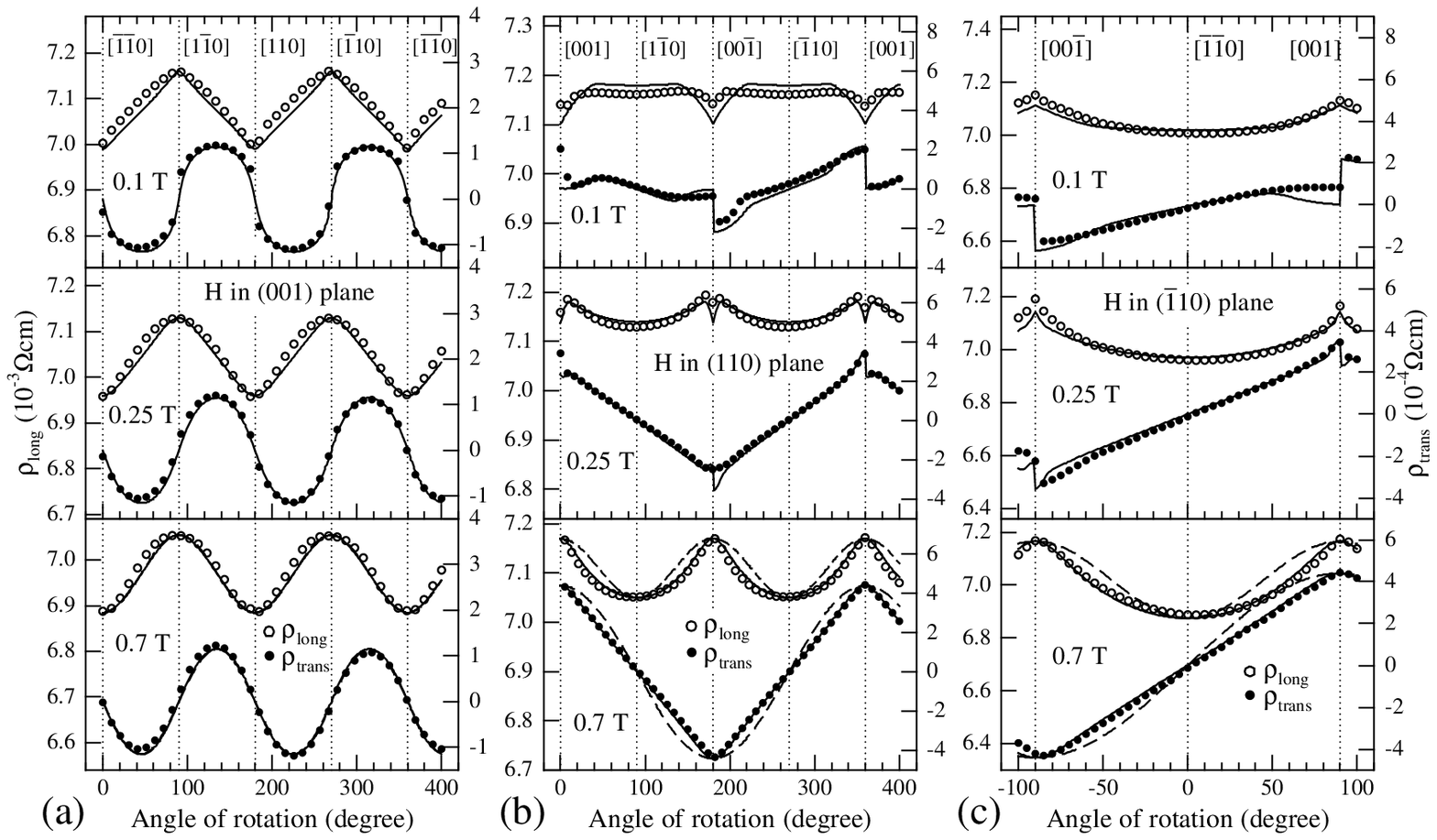}
\caption{\label{mt_001} Angle-dependent resistivities $\rho_{\mathrm{long}}$ (circles) and
$\rho_{\mathrm{trans}}$ (dots) of the (001) (Ga,Mn)As sample at 4.2 K. The measurements were carried
out at fixed field strengths of $\mu_0H$ = 0.1, 0.25, and 0.7 T with $\bm{H}$ rotated in (a) the (001),
(b) the (110), and (c) the ($\bar{1}$10) plane. The solid lines represent fits to the experimental data
using Eqs.~(\ref{rho001_110_tetragonal}) and one set of resistivity and anisotropy parameters. The dashed
lines at 0.7 T simulate the limiting case where $\bm{M}$ perfectly aligns with $\bm{H}$. In (a) the dashed
lines completely coincide with the solid lines.}
\end{figure*}

\begin{figure}
\includegraphics{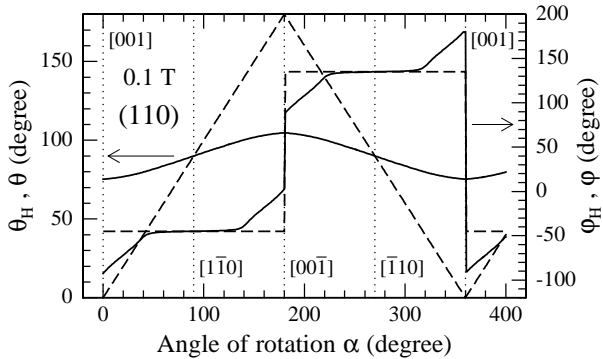}
\caption{\label{mag_110} Calculated polar (left axis) and azimuth (right axis) angles of the magnetic
field $\bm{H}$ (dashed lines) and the magnetization $\bm{M}$ (solid lines) for $\bm{H}$ rotated in the
(110) plane of the (001) (Ga,Mn)As sample at 0.1~T. The resistivity and anisotropy parameters used in
the calculation were derived from a fit to the angle-dependent resistivities shown in Fig.~\ref{mt_001}.}
\end{figure}

\subsection{(001) orientation}

The anisotropy of the (001)-oriented sample was probed by rotating $\bm{H}$ within the (001), (110),
and ($\bar{1}$10) planes. The corresponding angular dependences of $\rho_{\mathrm{long}}$ and
$\rho_{\mathrm{trans}}$, measured with the current direction along [110], are shown in Fig.~\ref{mt_001}.
At 0.7 T the Zeeman energy dominates the free enthalpy and MA only plays a minor role. As a consequence,
$\bm{M}$ is expected to nearly align with the applied magnetic field and to continuously follow the
motion of $\bm{H}$. In fact, the curves of $\rho_{\mathrm{long}}$ and $\rho_{\mathrm{trans}}$ at 0.7 T
are smooth and largely reflect the anisotropy of the resistivity tensor. With decreasing magnetic field
the influence of the MA increases and the orientation of $\bm{M}$ deviates more and more from the field
direction. Accordingly, jumps and kinks occur in the curves at 0.25 and 0.1 T, arising from sudden
movements of $\bm{M}$ caused by discontinuous displacements of the minimum of the free enthalpy. The
observed angular dependences of $\rho_{\mathrm{long}}$ and $\rho_{\mathrm{trans}}$ can be understood in
great detail by modeling the measured curves within the theoretical framework presented in Section III.
For this purpose, the resistivity and anisotropy parameters from Eqs.~(\ref{resistivity_parameters_1}),
(\ref{resistivity_parameters_2}), and (\ref{FE_001}) were determined by an iterative fit procedure.
Starting with an initial guess for the anisotropy parameters, the resistivity parameters were obtained
by fitting Eqs.~(\ref{rho001_110_tetragonal}) to the experimental data recorded at 0.7~T. Then the
anisotropy parameters were modified for an optimal agreement at 0.25 and 0.1~T, and the whole procedure
was repeated until no further improvement of the fit could be achieved. The unit vector $\bm{m}$ at any
given magnetic field $\bm{H}$ was calculated by numerically minimizing $G^{001}_{M}$ in Eq.~(\ref{FE_001})
with respect to the polar and azimuth angles of $\bm{M}$. With the exception of $A$, the resistivity
parameters turned out to be field independent within the accuracy of the fit and are given by
$B$ = $-2.3\times 10^{-4}$ $\Omega$~cm,
$C+c$ = $-1.7\times 10^{-4}$ $\Omega$~cm,
$b_1$ = $0.9\times 10^{-4}$ $\Omega$~cm, and
$D+d$ = $-4.4\times 10^{-4}$ $\Omega$~cm.
The resistivity parameter $A$ was found to decrease from $7.21\times 10^{-3}$ $\Omega$~cm at 0.1~T
to $7.08\times 10^{-3}$ $\Omega$~cm at 0.7~T, reflecting the negative-magnetoresistance behavior of
$\rho_{\mathrm{long}}$.\cite{Ohn98,Mat04} For the anisotropy parameters we obtained the values
$B_{c\parallel}$ = $-0.015$~T, $B_{c\perp}$ = 0~T, $B_{001}$ = 0.17~T, and $B_{\bar{1}10}$ = 0.002~T.
The theoretical curves calculated with these parameters are in excellent agreement with the
experiment and are drawn as solid lines in Fig.~\ref{mt_001}.

Once the anisotropy parameters are known, the orientations of the easy axes can be determined by
minimizing $G^{001}_{M}$ with respect to $\bm{m}$ at zero magnetic field. The easy axes are found
to lie within the (001) layer plane ($\theta$ = 90$^{\circ}$) at the azimuth angles $\varphi_1$ =
1.9$^{\circ}$ and $\varphi_2$ = 88.1$^{\circ}$ (see Fig.~\ref{coordinates}). The slight deviation
from the cubic [100] and [010] axes towards the [110] direction arises from the positive value of
$B_{\bar{1}10}$.

Using the resistivity and anisotropy parameters given above, plots similar to Fig.~\ref{Sim_mag} can
be drawn for each configuration and field strength, revealing in great detail the motion of $\bm{M}$.
Figure~\ref{mag_110} shows as an example the polar ($\theta_H$,$\theta$) and azimuth
($\varphi_H$,$\varphi$) angles of $\bm{H}$ and $\bm{M}$, respectively, plotted as a function of the
angle of rotation $\alpha$ for $\bm{H}$ rotated in the (110) plane at 0.1~T [see Fig.~\ref{mt_001}(b)].
While $\theta_H$ passes through all values between $0^{\circ}$ and $180^{\circ}$ (dashed line, left
axis), $\bm{M}$ remains very close to the (001) plane with $75^{\circ} < \theta < 105^{\circ}$ (solid
line, left axis). For $40^{\circ} < \alpha < 140^{\circ}$ and $220^{\circ} < \alpha < 320^{\circ}$,
where $\bm{H}$ is closer to the (001) plane than to the [001] axis, the azimuth angles $\varphi_H$
and $\varphi$ of $\bm{H}$ and $\bm{M}$, respectively, almost perfectly coincide. When $\bm{H}$
approaches the hard [001] axis, however, $\bm{M}$ tends towards the easy [100] axis (azimuth angles
$0^{\circ}$ and $180^{\circ}$) or to the easy [010] axis (azimuth angles $-90^{\circ}$ and
$90^{\circ}$). At $\alpha$ = $180^{\circ}$ and $\alpha$ = $360^{\circ}$, where $\bm{H}$ exactly
aligns with the [00$\bar{1}$] and [001] directions, respectively, the azimuth angle of $\bm{M}$
undergoes a sudden jump by $90^{\circ}$.

Considerable information can also be obtained by comparing the measured angular dependences with
those expected for the limiting case where $\bm{M}$ perfectly aligns with $\bm{H}$. To this end,
$\rho_{\mathrm{long}}$ and $\rho_{\mathrm{trans}}$ were calculated with $\bm{m}$ replaced by the
vector $\bm{h}=\bm{H}/H$ in Eqs.~(\ref{rho001_110_tetragonal}). The resulting curves are depicted by
the dashed lines in Fig.~\ref{mt_001}. For $\bm{H}$ rotated within the (001) plane [Fig.~\ref{mt_001}(a)],
the linear and quadratic $(\bm{n}\cdot \bm{m})$ terms in Eqs.~(\ref{rho001_110_tetragonal}) vanish and
we obtain the well-known $\cos^2\phi_j$ and $\cos\phi_j\sin\phi_j$ dependences of $\rho_{\mathrm{long}}$
and $\rho_{\mathrm{trans}}$, respectively. At 0.7~T the dashed curves coincide with the solid curves,
meaning that for $\bm{H}$ rotated within the layer plane the magnetization almost perfectly follows
the motion of the magnetic field. For lower fields, $\bm{M}$ remains in the layer plane since [001]
is a hard axis, but it increasingly deviates from $\bm{H}$ towards the easy [100] and [010] axes.
At 0.1~T it abruptly switches whenever $\bm{H}$ approaches the somewhat harder [110] and [$\bar{1}$10]
axes, leading to the kinks observed for $\rho_{\mathrm{long}}$.

The rotation of $\bm{H}$ within a plane perpendicular to the layer is accompanied by significant
differences in the orientations of $\bm{H}$ and $\bm{M}$, even for 0.7 T. This is clearly demonstrated
in Figs.~\ref{mt_001}(b) and (c), where the dashed curves represent the
$(\bm{n}\cdot \bm{m})^2 = \cos^2\phi_n$ and $(\bm{n}\cdot \bm{m})=\cos\phi_n$ dependences of
$\rho_{\mathrm{long}}$ and $\rho_{\mathrm{trans}}$ in Eqs.~(\ref{rho001_110_tetragonal}), respectively,
with $\phi_n$ denoting the angle between $\bm{m}$ and $\bm{n}$. At 0.7~T, $\bm{H}$ and $\bm{M}$ coincide
whenever $\bm{H}$ is orientated parallel or perpendicular to the layer plane. At lower fields this is
no longer true and $\bm{M}$ avoids the perpendicular direction by tending towards the easy [100] and
[010] axes (see Fig.~\ref{mag_110}). Accordingly, the differences between the minimum and maximum values
of $\rho_{\mathrm{long}}$ and $\rho_{\mathrm{trans}}$ are drastically reduced at 0.25 and 0.1~T.

It should be emphasized that the magnetotransport measurements give clear evidence for the tetragonal
distortion of the (Ga,Mn)As layer: First, the parameters $B_{c\parallel}$ and $B_{c\perp}$, representing
the in-plane and out-of-plane contributions to the cubic-anisotropy term in $G^{001}_{M}$, significantly
differ. Second, Eqs.~(\ref{rho001_110_cubic}), which have been derived for the case of perfect cubic
symmetry, correctly reproduce the measured amplitudes of $\rho_{\mathrm{long}}$ for the in-plane
configuration in Fig.~\ref{mt_001}(a), but do not so for the two out-of-plane configurations in
Figs.~\ref{mt_001}(b) and (c). Moreover, it is worth noting that the fits presented in Fig.~\ref{mt_001}
could be improved even further by taking into account higher-order terms in the series expansion of the
resistivity tensor in Eq.~(\ref{series_expansion}). However, since the agreement achieved in second order
is more than satisfactory and since the mathematical expressions for the resistivities would become much
more complicated, higher-order terms have not been considered in the present study.

\begin{figure}
\includegraphics{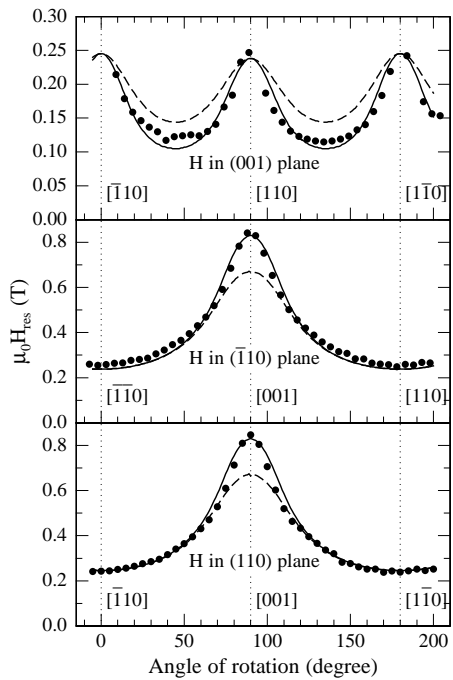}
\caption{\label{fmr_001} Angle-dependent FMR fields of the (001) (Ga,Mn)As sample at 5 K for $\bm{H}$
rotated in the (001), ($\bar{1}$10), and (110) planes. The solid lines represent the result of a least
squares fit, the dashed lines were calculated using $g$ = 2.0 and the anisotropy parameters estimated
from the magnetotransport data.}
\end{figure}

The results of FMR measurements are presented in Fig.~\ref{fmr_001}. It shows the measured and simulated
angular dependences of the resonance field $H_{\mathrm{res}}$ for $\bm{H}$ rotated within the (001),
($\bar{1}$10), and (110) planes. The dashed lines, reproducing only roughly the experimental curves, were
numerically calculated using Eq.~(\ref{FE_001}), Eq.~(\ref{resonance_condition}), $g$ = 2.0, and the
anisotropy parameters derived from the magnetotransport measurements. The agreement between experiment
and theory is significantly improved using $g$ = 1.9 and the slightly higher values $B_{c\parallel}$ =
-0.02~T, $B_{c\perp}$ = 0~T, $B_{001}$ = 0.24~T, and $B_{\bar{1}10}$ = 0.002~T, which were obtained by a
least squares fit based on Eqs.~(\ref{FE_001}) and (\ref{resonance_condition}). The values agree within
30\% with the anisotropy parameters determined from magnetotransport. The reason for the remaining
difference between the two sets of parameters is not yet clear. Inevitable sample heating up to
150$^\circ$C for less than 30 min during the Hall-bar preparation as well as the different lateral sizes
of our samples (shape anisotropy) are not expected to account for the observed variation.

\begin{figure}
\includegraphics{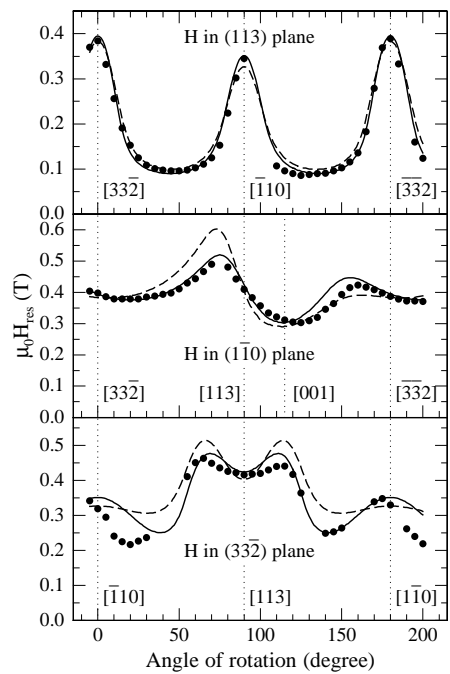}
\caption{\label{fmr_113} Angle-dependent FMR fields of the (113)A (Ga,Mn)As sample at 5 K for $\bm{H}$
rotated in the (33$\bar{2}$), (1$\bar{1}$0), and (113)A planes. The solid and dashed lines represent the
results of least squares fits with and without considering a uniaxial term along [001] in $G_{M}^{113}$,
respectively.}
\end{figure}

\begin{figure*}
\includegraphics{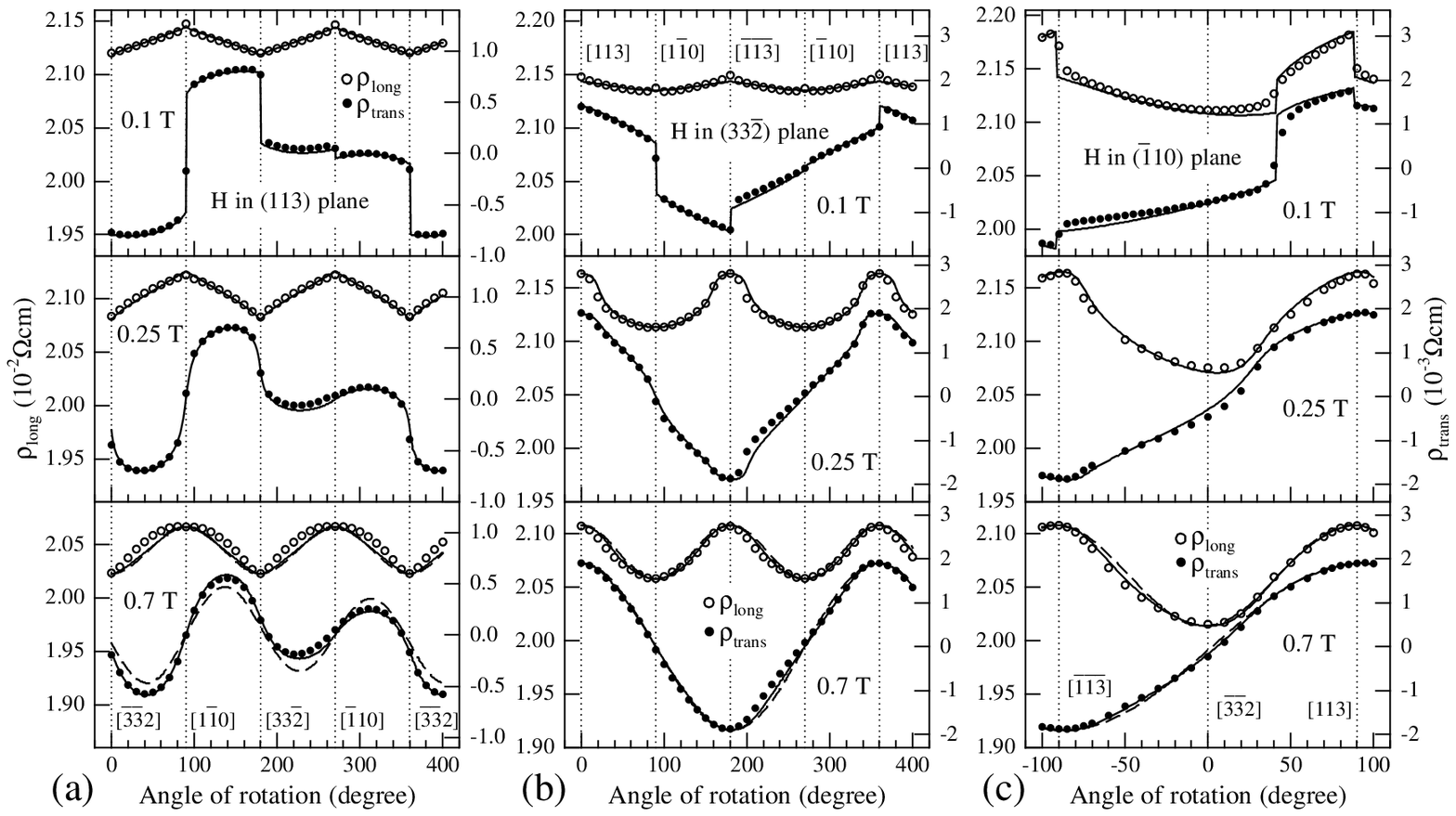}
\caption{\label{mt_113} Angle-dependent resistivities $\rho_{\mathrm{long}}$ (circles) and
$\rho_{\mathrm{trans}}$ (dots) of the (113)A (Ga,Mn)As sample at 4.2 K. The measurements were carried
out at fixed field strengths of $\mu_0H$ = 0.1, 0.25, and 0.7 T with $\bm{H}$ rotated in (a) the (113),
(b) the (33$\bar{2}$), and (c) the ($\bar{1}$10) plane. The solid lines represent fits to the experimental
data using Eqs.~(\ref{rho113_33m2_tetragonal}) and one set of resistivity and anisotropy parameters. The
dashed lines at 0.7 T simulate the limiting case where $\bm{M}$ perfectly aligns along $\bm{H}$.}
\end{figure*}

\subsection{(113)A orientation}

As already mentioned in Section III B, our experimental data suggest the existence of a lattice
distortion along [001] even in the (Ga,Mn)As film grown on GaAs(113)A substrate. This is demonstrated
in Fig.~\ref{fmr_113}, which shows the measured and simulated angular dependences of the FMR field
$H_{\mathrm{res}}$ for $\bm{H}$ rotated within the (33$\bar{2}$), (1$\bar{1}0)$, and (113) planes.
The solid and dashed lines depict the results of least squares fits using Eqs.~(\ref{FE_113}) and
(\ref{resonance_condition}) with and without considering the uniaxial term $B_{001} m_z^2$, respectively.
Even though no perfect simulation of the measured curves could be achieved, the solid curve is much
closer to the experimental data than the dashed one. Optimal agreement is obtained for $g=2.0$ and the
values $B_{c\parallel}$ = $B_{c\perp}$ = -0.046~T, $B_{113}$ = 0.032~T, and $B_{001}$ = 0.053~T of the
anisotropy parameters (solid line in Fig.~\ref{fmr_113}). Similar to the case of the (001) layer, the
uniaxial in-plane contribution is almost negligible with $B_{\bar{1}10}$ = -0.005~T.

The magnetotransport data, measured with the current direction along [33$\bar{2}$], are depicted in
Fig.~\ref{mt_113}. The figure shows the experimental and simulated angular dependences of
$\rho_{\mathrm{long}}$ and $\rho_{\mathrm{trans}}$ for $\bm{H}$ rotated within the (113), (33$\bar{2}$),
and ($\bar{1}$10) planes. Using the fit procedure described in Section IV A, the values of the
resistivity parameters in Eqs.~(\ref{rho113_33m2_tetragonal}) are obtained as
$B$ = $-8.1\times 10^{-4}$ $\Omega$~cm,
$C$ = $-8.7\times 10^{-4}$ $\Omega$~cm,
$b$ = $7.2\times 10^{-3}$ $\Omega$~cm,
$c$ = $5.1\times 10^{-4}$ $\Omega$~cm,
$D$ = $-2.1\times 10^{-3}$ $\Omega$~cm, and
$d$ = $2.1\times 10^{-4}$ $\Omega$~cm.
The sum $A+2a/11$ decreases from $21.61\times 10^{-3}$ $\Omega$~cm at 0.1~T to $20.79\times 10^{-3}$
$\Omega$~cm at 0.7~T. For the anisotropy parameters we obtained the values $B_{c\parallel}$ = -0.046~T,
$B_{c\perp}$ = -0.03~T, $B_{113}$ = 0.018~T, $B_{001}$ = 0.02~T, and $B_{\bar{1}10}$ = -0.008~T.
Similar to the (001) sample, they are smaller than the values determined by the FMR study, however,
the discrepancy is less than a factor of about two. The calculated curves, represented by the solid
lines in Fig.~\ref{mt_113}, are in excellent agreement with the measured data. Again, the dashed
curves simulate the case of a magnetization which perfectly aligns with $\bm{H}$. A comparison between
the dashed and solid curves reveals that the magnetization at 0.7~T almost perfectly follows the motion
of the magnetic field in the two out-of-plane configurations [Figs.~\ref{mt_113}(b) and (c)], whereas for
$\bm{H}$ rotated within the layer plane [Fig.~\ref{mt_113}(a)] $\bm{M}$ significantly deviates from
$\bm{H}$. According to the model calculations, the latter behavior arises from the cubic terms and the
uniaxial [001] contribution in $G^{113}_{M}$, resulting in a deflection of $\bm{M}$ towards the (001)
plane. The asymmetry of $\rho_{\mathrm{trans}}$ in Fig.~\ref{mt_113}(a) partly results from this
deflection and partly from the last term in Eqs.~(\ref{rho113_33m2_tetragonal}) which originates from
the tetragonal distortion. A similar asymmetry has been observed by Muduli et al.\cite{Mud05} in
Fe$_3$Si films grown on GaAs(113)A substrates. There, the asymmetry has been explained by third-order
terms in $\rho_{\mathrm{trans}}$. We cannot rule out that in the (113)A (Ga,Mn)As sample under study
higher-order terms contribute to the asymmetry, too. However, since the present model, including terms
up to second order, fully accounts for the observed angular dependence of $\rho_{\mathrm{long}}$ and
$\rho_{\mathrm{trans}}$, it has not been considered as mandatory to include them.

Using the anisotropy parameters obtained from the curve fits, the orientations of the easy axes were
determined by minimizing $G^{113}_{M}$ with respect to $\bm{m}$ at zero magnetic field. We find the
easy axes in the (113)A layer at the angles (see Fig.~\ref{coordinates}) $\theta_1$ = 92$^{\circ}$,
$\varphi_1$ = -3.4$^{\circ}$, and $\theta_2$ = 92$^{\circ}$, $\varphi_2$ = 93.4$^{\circ}$, i.e.,
very close to the [100] and [010] axes, in qualitative agreement with the results presented in
Refs.~\onlinecite{Bih06} and \onlinecite{Wan05c}.

\section{Summary}

A series expansion of the resistivity tensor with respect to the magnetization components yields
general expressions for the longitudinal and transverse resistivites in single-crystalline ferromagnets
with cubic and tetragonal symmetry. The expressions, applicable to (Ga,Mn)As layers with arbitrary surface
index, were used to quantitatively model the angular dependences of the resistivities, measured in (001)
and (113)A (Ga,Mn)As films as a function of magnetic field orientation. Whereas the curves at 0.7 T
largely reflect the anisotropy of the resistivity tensor, the curves at 0.25 and 0.1 T are strongly
affected by magnetic anisotropy, allowing access to anisotropy parameters. The magnetotransport data and
comparative ferromagnetic resonance studies reveal an inclined uniaxial anisotropy along [001] in the
(113)A-oriented (Ga,Mn)As layers in addition to the usual in- and out-of-plane contributions known from
(001) layers.

\begin{acknowledgments}
This work was supported by the Deutsche Forschungsgemeinschaft (Li 988/4 and SFB 631).
\end{acknowledgments}


\begin{thebibliography}{99}
\bibitem{comment}
In most publications the energy of the magnetic system is referred to
as free energy. Since the independent variable is represented by the magnetic field
$\bm{H}$ and not by the magnetization $\bm{M}$, this energy should be
correctly denoted as free enthalpy.
\end{thebibliography}
\end{document}